\documentclass[epj,referee]{svjour}
\usepackage{graphics}
\begin{document}
\title{Non-Markovian entanglement dynamics  in coupled superconducting qubit systems}
\author{Wei Cui\inst{1,2} \and Zai-Rong Xi\inst{1} 
\thanks{\emph{Present address:} zrxi@iss.ac.cn}%
\and Yu Pan\inst{1,2}}                     
%
%
\institute{Key Laboratory of Systems and Control, Institute of
Systems Science, Academy of Mathematics and Systems Science, Chinese
Academy of Sciences, Beijing 100190, P.~R.~China \and Graduate
University of Chinese Academy of Sciences, Beijing 100039,
P.~R.~China}
\date{Received: date / Revised version: date}
%
\abstract{ We theoretically analyze the entanglement generation and
dynamics by coupled Josephson junction qubits. Considering a
current-biased Josephson junction (CBJJ),
  we generate  maximally entangled states. In particular, the entanglement
  dynamics is considered as a function of the decoherence
parameters, such as the temperature, the ratio
$r\equiv\omega_c/\omega_0$ between the reservoir cutoff frequency
$\omega_c$ and the system oscillator frequency $\omega_0$,
 and the energy levels split of the superconducting circuits in the
non-Markovian master equation. We analyzed the entanglement sudden
death (ESD) and entanglement sudden birth (ESB) by the non-Markovian
master equation. Furthermore, we find that the larger the ratio $r$
and the thermal energy $k_BT$, the shorter the decoherence. In this
superconducting qubit system we find that the entanglement can be
controlled and the ESD time can be prolonged by adjusting the
temperature and the superconducting phases $\Phi_k$ which split the
energy levels.
} 
\maketitle
\section{Introduction}
Entanglement is one of the remarkable features of quantum mechanics.
Briefly, entanglement refers to correlated behavior of two or more
particles that cannot be described classically, the properties of
one particle can depend on those of another (typically distant)
particle in a way that only quantum mechanics can explain.
 Einstein-Podolsky-Rosen (EPR) entangled states, probably the
 simplest and most interesting entangled states, have been employed
 not only to test Bell's inequality, but
 also to realize quantum cryptography, quantum teleportation, and
 quantum computation \cite{Stolze:08,Bellac:06,Mermin:07,Rowe:01,Nori:09}. It has become clear that
 entanglement is
 a new resource for tasks that cannot be performed by means of
 classical resources \cite{Bennett,Zoller}. Although many results
have been obtained (for example, see the review papers
\cite{Amico,Horodecki}), the theory of quantum entanglement has open
questions, like (i) how to optimally detect entanglement
theoretically and practically; (ii) how  to reverse the inevitable
process of degradation of entanglement; and (iii) how to
characterize, quantify and control entanglement. The focus of this
paper is a theory study of entanglement dynamics and control in a
coupled Josephson junction system.

Superconducting quantum circuits \cite{You2} are the subject of
intense research at present. Josephson devices can serve as quantum
bits (qubits) in quantum information and that quantum logic
operations could be performed by controlling gate voltages or
magnetic fields \cite{You2}. Moreover, for its scalable and
macroscopic property, Josephson junctions offer one of the most
promising candidate served as hardware implementation of quantum
computers \cite{Berkley,McDermott,Liu,Dimitris,Wei,Schulz,You}. The
charge, flux, and phase qubits are three basic types of
superconducting qubits \cite{You2} depending on which dynamical
variable is most well defined.  Operations with multiple
superconducting qubits have also been performed. Several types of
Josephson-junction qubits have been proposed and explored in the
laboratory. The first solid-state quantum gate has been demonstrated
with charge qubits \cite{Yamamoto}. For flux qubits, two-qubit
coupling and a controllable coupling mechanism have been realized
\cite{Hime,Niskanen}.  However, due to the difficulty to decouple
the qubits from the environments in solid-state systems, only the
two-qubit entanglement of the superconducting qubits has been
observed in experiments \cite{Steffen}, while the entangled states
up to eight \cite{Haffner} or six photonic \cite{Pan} qubits have
been experimentally reported. Therefore, the future generation of
multi-particle entangled states  for solid state quantum computation
would be a significant,  step towards quantum information
processing. A challenge is quantum decoherence, because any pure
quantum state used evolves into a mixed state due to the unavoidable
interactions with the environment. Decoherence describes the
environment-induced suppression of the quantum mechanical coherence
properties and interference ability, which transforms the quantum
system into classical one. The description of this process requires
us to take into account not only the degrees of freedom of the
system of interest, but also those of the environment. Decoherence
of the Josephson junction qubits is considered to be the major
impediment for quantum logic gate operations. Thus, short coherence
times limit both the manipulation of the qubit state and information
storage. In all superconducting qubits, both the spectrum of charge
noise and the critical current fluctuations as it is display a $1/f$
behavior at low frequencies. Moreover, both charge noise and
critical current fluctuations can be phenomenologically explained by
modeling the environment as a collection of discrete bistable
fluctuators, representing charged impurities hopping between
different locations in the substrate or in the tunnel barrier. Many
efforts \cite{Robert} searched for decoherence mechanisms that
suppress the decoherence time.

On the path to quantum computing, superconducting qubits \cite{You2}
are clearly among the most promising candidates. References
\cite{Ashhab1,Ashhab2,Ashhab3} showed that such two-level systems
can themselves be used as qubits, allowing for a well controlled
initialization, universal sets of quantum gates, and readout. Thus,
a single current-biased Josephson junction can be considered as a
multi-qubit register. It can be coupled to other junctions to allow
the application of quantum gates to an arbitrary pair of qubits in
the system. These results \cite{Ashhab1,Ashhab2,Ashhab3} indicate an
alternative way to control qubits  coupled to naturally formed
quantum two-level systems, for improved superconducting quantum
information processing. Nevertheless, the path is long, and there
are quantitative technological obstacles to be overcome, notably
increasing the decoherence time, improving the fidelity of the
read-out, and strengthening the entanglement distillation in its
dynamics.

Within the theory of open quantum system
\cite{Onofrio:93,Onofrio:96,Breuer} the dissipative dynamics can be
described by the master equation of the reduced density matrix. In
general this equation is obtained by tracing over the environment
variables, after performing a series of approximations. The
Born-Markovian approximation is usually used in deducing the master
equation, which neglects the correlations between the system and the
reservoir.  The Markovian approximation leads to a master equation
which can be cast into the so called Lindblad form. Master equations
in the Lindblad form are characterized by the fact that the
dynamical group of the system satisfies both the semigroup property
and the complete positivity condition.  However, in the
superconducting  qubit system the Markovian approximation is not
justified \cite{Chirolli,JSZhang}. In recent years, non-Markovian
quantum dissipative systems
\cite{Cui1,YJZhang,Lajos,Suominen,Gambetta,Vacchini,Wolf,Burkard,Breuer:09,Piilo:09}
have attracted much attention due to its fundamental importance in
quantum information processing .

In this paper, we will  focus on the dynamics of the generated
superconducting entangled state in a non-Markovian environment by
using the master equation method. It determines how much quantum
information can be reliably transmitted over the noisy quantum
channels. We then discuss schemes for controlling the entanglement
between the superconducting quantum circuits. Nowadays,
quantum-state engineering, i.e., active control over the coherent
dynamics of suitable quantum-mechanical systems, has become a
fascinating prospect of modern physics
\cite{Ashhab1,Ashhab2,Ashhab3}. Quantum decoherence and entanglement
control pave the way for future long coherent time quantum
information processing and computation.

\begin{figure*}
\centerline{\scalebox{0.6}[0.6]{\includegraphics{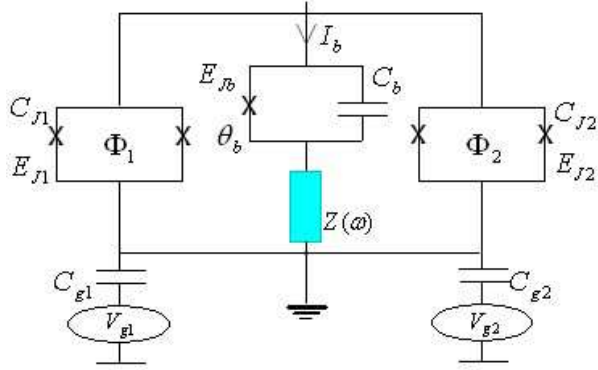}}}
\caption{Color online) Schematic diagram of the coupled-qubit
circuit with a biased-current source of impedance $Z(\omega)$. Two
Josephson charge   qubits are controllably coupled to a common
current-biased Josephson junction, which operates as a Josephson
phase qubit and acts as a coupler. Two Josephson qubits and
current-biased Josephson junction with the electromagnetic
environment represented by the impedance $Z(\omega)$.}
\end{figure*}

This paper is organized as follows. We first introduce
superconducting qubits decoherence and the quantum non-Markovian
master equation for driven open quantum systems. In Sec. II, we
shall briefly recall the physics of the Cooper pair box, and give
the model of the superconducting qubits interacting with the  bath.
In Sec. III, we consider the entanglement dynamics by the
non-Markovian master equation. Both entanglement sudden death (ESD)
and entanglement sudden birth (ESB) are analyzed by numerical
simulation.
Conclusions  are given in Sec. IV.

\section{The model}
In this work, we present an experimentally implementable method to
couple two Josephson charge qubits and to generate, detect and
control of macroscopic quantum entangled states in this charge-qubit
system.
 Let us study the superconducting
circuit shown in Fig.1, where two single Cooper pair boxes (CPBs)
are connected via a common bus, i.e., a current-biased Josephson
junction (CBJJ). Each qubit consists of a gate electrode of
capacitance $C_g$ and a single Cooper pair box with two ultrasmall
identical Josephson junctions of capacitance $C_J$ and Josephson
energy $E_J$, forming a superconducting quantum interference device
(SQUID) ring threaded by a flux $\Phi$ and with a gate voltage $V$.
The superconducting phase difference across the $k$th qubit is
represented by $\Phi_k,~~k=1,2$. The large CBJJ has capacitance
$C_b$, phase drop $\theta_b$, Josephson energy $E_b$, and a bias
current $I_b$. The reason for this choice is that this circuit can
be easily generalized to include more qubits, coupled by a common
CBJJ. For the detail of controllable coupling of superconducting
qubits, one can refer to
\cite{You:031,You:032,You:05,Wei:05,Wei:06,Wei:062,Ashhab:06,Ashhab:07,Liu:07}.The
Hamiltonian without dissipation can be written as
 \begin{equation}
 \label{Hamiltonian}
\hat{H}_s=\sum_{k=1}^24E_c(\hat{n}_k-n_{gk})^2-E_J(\Phi_k)\cos\hat{\theta}_k+\hat{H}_{kb}+\hat{H}_b,
 \end{equation}
where
\begin{eqnarray}
\hat{H}_{kb}&=&\pi
C_gE_J(\Phi_k)\hat{\theta}_b\sin\hat{\theta}_k/\left[\Phi_0(C_g+2C_J^0)\right],\nonumber\\
\hat{H}_b&=&\hat{Q}_b^2/(2\tilde{C}_b)-E_{Jb}\cos\hat{\theta}_b-\Phi_0I_b/(2\pi\hat{\theta}_b).\nonumber
 \end{eqnarray}
Here $E_c=e^2/[2(C_g+2C_J^0)]$ is the single-electron charging
energy of a single CPB. For simplicity, we assume that $E_c$ and
$E_J^0$ are the same for the two CPBs. $C_g$ and $C_g^0$ are the
capacitances of the gate electrode and identical Josephson
junctions, respectively. The Josephson energy $E_J(\Phi_k)$ of the
$k$th dc SQUID is $E_J(\Phi_k)=2E_J^0\cos(\pi\Phi_k/\Phi_0)$, where
$E_J^0$ represents the Josephson energy of a single Josephson
junction, $\Phi_k$ denotes the external flux piercing the SQUID loop
of the $j$-th CPB, and $\Phi_0$ is the flux quantum.
$\hat{Q}_b=2\pi\hat{p}_b/\Phi_0$ is the operator of charges on the
CBJJ. $E_{Jb}$ and
$\tilde{C}_b=C_{Jb}+\sum_{k=1}^2(C_{Jk}^{-1}+C_{gk}^{-1})^{-1}=C_{Jb}+2((2C_J^0)^{-1}+C_g^{-1})^{-1}$
are the Josephson energy and effective capacitance of the CBJJ,
respectively. Also, $\Phi_0=h/(2e)$ is the flux quantum. The
operators $\hat{n}_k$ and $\hat{\theta}_k$ satisfy the commutation
relations $[\hat{\theta}_k ,\hat{n}_k]=i$ (we assume $\hbar=1$),
which describe the excess number of Cooper pairs and the effective
phase across the junctions in the $k$-th CPB, respectively. In
addition, the phase operator $\hat{\theta}_b$ for the CBJJ and its
conjugate $\hat{p}_b$ satisfy another commutation relation
$[\hat{\theta}_b, \hat{p}_b]=i$.

Suppose that the CPBs are biased at the charge degenerate point,
such that $n_{gk}=C_{gk}V_k/(2e)=1/2$ (when $V_k=e/C_{gk}$). The two
energy levels of the $k$-th CPB corresponding to $n_k=0,1$ are close
to each other and far separated from other high-energy levels. In
this case, they behave as effective two level systems (with the
basis
$\left\{|\uparrow_k\rangle=|n_k=0\rangle,~~|\downarrow_k\rangle=|n_k=1\rangle\right\}$)
\cite{Wei1}. It is well known that the CBJJ can be approximated as a
harmonic oscillator \cite{Wei1,Wei,Zhang1,Zhang2}, if it is biased
as $I_b\ll I_0=2\pi E_{Jb}/{\Phi_0}$. Here, we consider a very
different case, i.e., the biased dc current $I_b$ is slightly
smaller than the critical current $I_0$, and thus the CBJJ has only
a few bound states. The two lowest energy states $|0_b\rangle$ and
$|1_b\rangle$ are selected to define a Josephson phase qubit acting
as a two-level date bus. Under such condition, the Hamiltonian of
the CBJJ reduces to $\hat{H}_b=\hbar\omega_b\hat{\sigma}_b^z$, with
$\hat{\sigma}_b^z=|0_b\rangle\langle0_b|-|1_b\rangle\langle1_b|$
being the standard Pauli operator and $\omega_b=E_{b1}-E_{b0}$ the
eigenfrequency.

Under the rotating-wave approximation, the Hamiltonian $H$ can be
rewritten as
\begin{eqnarray}
 \label{Hamiltonian2}
H_s&=&\sum_{k=1}^2\left(\frac{E_{Jk}(\Phi_k)}{2}\sigma_z^{(k)}-\frac{E_{Ck}(n_{gk})}{2}\sigma_x^{(k)}\right)+\omega_b\sigma_z^{(b)}
\nonumber\\&&+J\left(\sigma_+^1\sigma_-^2+\sigma_+^2\sigma_-^1\right),
 \end{eqnarray}
with
\begin{equation}
 \begin{array}{rcl}
\sigma_x^{(k)}&=&|+\rangle_{kk}\langle-|+|-\rangle_{kk}\langle+|,\\
\sigma_y^{(k)}&=&-i|+\rangle_{kk}\langle-|+i|-\rangle_{kk}\langle+|,\\
\sigma_z^{(k)}&=&|+\rangle_{kk}\langle+|-|-\rangle_{kk}\langle-|,
 \end{array}\nonumber
 \end{equation}
 and $\sigma_{\pm}^{(k)}=\left(\sigma_x^{(k)}\pm i\sigma_y^{(k)}\right)/2$, where
$|+\rangle_k=(|0\rangle_k+|1\rangle_k)/\sqrt{2}$ and
$|-\rangle_k=(|0\rangle_k-|1\rangle)/\sqrt{2}$. Note that, the
charging energy $E_{Ck}(n_{gk})=2e^2(1-2n_{gk})/C_k$ of the $k$-th
Josephson charge qubit  can be switched off by setting the gate
voltage $V_k$ such that $n_{gk}=1/2$. Also, by adjusting the
external flux $\Phi_k$, the Josephson energy of the $k$-th qubit
$E_{Jk}(\Phi_k)=2E^0_J\cos(\pi\Phi_k/\Phi_0)$ can be set to the
strongest coupling $(\Phi_k=0)$ and the decoupling
$(\Phi_k=\Phi_0/2)$. This achieves the controllability of the
present quantum circuit. The coefficient $J$ is the coupling
strength between the $k$-th qubit and Josephson junction  with
$J=\lambda_1\lambda_2/E_{c},~~\lambda_k=2E_J^0\sin(\Phi_k/2)$.

Obviously, if we assume   the two qubits are both at the charge
degenerate point $n_{gk}=1/2$, and $\Phi_k=\Phi_0/2$ with the
separated initial state $|g\rangle_b|g\rangle_1|e\rangle_2$, the
evolution operator of the corresponding two qubit system is given by
\begin{equation}
 \label{operator}
U(t)=\exp\left[-iJt(\sigma_+^1\sigma_-^2+\sigma_+^2\sigma_-^1)-i\omega_bt\sigma_z^{(b)}\right],
 \end{equation}
and the system's evolution is given by

\begin{eqnarray}
U(t)|g\rangle_b|g\rangle_1|e\rangle_2=e^{-i\omega_bt\sigma_z^{(b)}}|g\rangle_be^{-iJt(\sigma_+^1\sigma_-^2+\sigma_+^2\sigma_-^1)}|g\rangle_1|e\rangle_2\nonumber\\
=(\cos\omega_bt+i\sin\omega_bt)|g\rangle_b\left[\cos
Jt|g\rangle_1|e\rangle_2-i\sin
Jt|e\rangle_1|g\rangle_2\right].\nonumber\\
 \end{eqnarray}
If we choose $Jt=\pi/4$, we can obtain the maximally entangled state
$|\Psi\rangle=(|g\rangle_1|e\rangle_2-i|e\rangle_1|g\rangle_2)/\sqrt{2}$.

\section{ Entanglement dynamics}

\subsection{Markovian and non-Markovian master equations}
We account for the dissipation due to electromagnetic fluctuations.
They can be modeled by an effective impedance $Z(\omega)$, placed in
series with the voltage source and producing a fluctuating voltage.
The impedance is embedded in the circuit shown in Fig.1, which
further modifies the spectrum of voltage fluctuations. In general,
the environment consists of a large set of harmonic oscillators,
each of which interacts weakly with the system of interest, i.e.
$H_{bath}=
\sum_i\left[\frac{p_i^2}{2m_i}+\frac{m_iw_i^2}{2}x_i^2\right] $. As
mentioned in Sec. 1, we will use the master equation method to study
the entanglement dynamics and control.

 The analysis of the time evolution of open quantum system plays an
important role in many applications of modern physics. With the
Born-Markovian approximation the dynamics is governed by a master
equation of relatively simple form \cite{Breuer}
\begin{equation}
\label{Master equation1} \frac{d}{d
t}\rho(t)=-i[\hat{H}_s(t),\rho(t)]+\sum_m \gamma_m\mathcal
{D}[C_m]\rho(t),
\end{equation}
with a time-independent generator in the Lindblad form.
 This is the most general form for the generator of a quantum
 dynamical semigroup. The Hamiltonian $\hat{H}_S(t)$ describes the coherent
 part of the time evolution. Non-negative quantities $\gamma_m$ play the role of relaxation rates
 for the different decay modes of the open system. The operators $C_m$ are usually
 referred to as Lindblad operators which represent the various
 decay modes, and the corresponding density matrix equation (\ref{Master equation1}) is
 called the Lindblad master equation. The solution of Eq. (\ref{Master
 equation1}) can be written in terms of a linear map $V(t)=\exp(\mathcal
 {L}t)$ that transforms the initial state $\rho(0)$ into the state
 $\rho(t)=V(t)\rho(0)$ at time $t$. The physical interpretation of
 this map $V(t)$ requires that it preserves the trace and the
 positivity of the density matrix $\rho(t)$.
The most important physical assumption which underlies  Eq.
(\ref{Master equation1}) is the validity of the Markovian
approximation of short environmental correlation times. With this
approximation, the environment acts as a sink for the system
information. Due to the system-reservoir interaction, the system of
interest loses information on its state into the environment, and
this lost information does not play any further role in the system
dynamics.

If the environment has a non-trivial structure, then the seemingly
lost information can return to the system at a later time leading to
non-Markovian dynamics with memory. This memory effect is the
essence of non-Markovian dynamics
\cite{Cui1,Lajos,Suominen,Gambetta,Vacchini,Wolf,Burkard,Breuer:09,Piilo:09},
which is characterized by pronounced memory effects, finite revival
times and non-exponential relaxation and decoherence. Non-Markovian
dynamics plays an important role in many fields of physics, such as
quantum optics, quantum information, quantum chemistry process,
especially in solid state physics \cite{Breuer:09}. As a consequence
the theoretical treatment of non-Markovian quantum dynamics is
extremely demanding. However, in order to take into account quantum
memory effects, an integro-differential equation is needed which has
complex mathematical structure, thus preventing generally to solve
the dynamics of the system of interest. An appropriate scheme is the
time-covolutionless (TCL) projection operator technique
\cite{Breuer,Breuer:09,Piilo:09} which leads to a time-local first
order differential equation for the density matrix.


By tracing out the bath degrees of freedom, we find for $\rho$ a
non-Markovian evolution equation
\begin{equation}
\label{Master equation2} \frac{d}{d
t}\rho(t)=-i[\hat{H}_s(t),\rho(t)]+\sum_m\Delta_m(t)\mathcal
{D}[C_m(t)]\rho(t).
\end{equation}
with the super-operator $\mathcal {D}[L]\rho=L\rho
L^{\dag}-\frac{1}{2}L^{\dag}L\rho-\frac{1}{2}\rho L^{\dag}L$. The
first term describes the unitary part of the evolution. The latter
involves a summation over the various decay channels labeled by $m$
with corresponding time-dependent decay rates $\Delta_m(t)$ and
arbitrary time-dependent system operators $C_m(t)$. In the simplest
case, the rates $\Delta_m$ as well as the Hamiltonian $\hat{H}_s$
and the operators $C_m$ are assumed to be time-independent, that is,
it is the Markovian case. Note that, for arbitrary time-dependent
operators $\hat{H}_s(t)$ and $C_m(t)$, and for $\Delta_m(t)\geq0$
the generator of the master equation (\ref{Master equation2}) is
still in Lindblad form at each fixed time $t$, which may be
considered as time-dependent quantum Markovian process. However, if
one or several of the $\Delta_m(t)$ become temporarily negative,
which expresses the presence of strong memory effects in the reduced
system dynamics, the process is then said to be non-Markovian.

%

For the system considered in Fig.1, two Josephson junction qubits
coupled by the common current-biased Josephson junction, the
bipartite dynamics is
\begin{equation}\label{master equation}
\frac{d\rho(t)}{dt}=-i[H_s,
\rho]+\sum_{k=1}^2(\Gamma_1D[\sigma_k^-]\rho+\Gamma_2D[\sigma_k^+]\rho),
\end{equation}
where $\Gamma_1=\Delta(t)+\gamma(t)$ and
$\Gamma_2=\Delta(t)-\gamma(t)$. The time dependent coefficients
$\Delta(t)$ and $\gamma(t)$ are diffusive term and damping term,
which can be written as follows
\begin{equation}
\nonumber
 \Delta(t)=\int_0^td\tau k(\tau)\cos(\omega\tau),
 \end{equation}
\begin{equation}
 \gamma(t)=\int_0^td\tau \mu(\tau)\sin(\omega\tau),
 \end{equation}
with
\begin{equation}
\nonumber
 k(\tau)=2\int_0^{\infty}d\omega
J(\omega)\coth[\hbar\omega/2k_BT]\cos(\omega \tau), \label{k}
 \end{equation}
\begin{equation}
\mu(\tau)=2\int_0^{\infty}d\omega J(\omega)\sin(\omega \tau),
\label{mu}
 \end{equation}
being the noise and the dissipation kernels, respectively. In this
paper we choose the Ohmic spectral density with a
 Lorentz-Drude cutoff function,
\begin{equation}
J(\omega)=\frac{2\gamma_0}{\pi}\omega\frac{\omega_c^2}{\omega_c^2+\omega^2},
 \end{equation}
 where $\gamma_0$ is the frequency-independent damping constant and usually assumed to be $1$.
 $\omega$ is the frequency of the bath, and $\omega_c$ is the high-frequency
 cutoff. The analytic expression for the coefficients $\gamma(t)$and $\Delta(t)$ are given in \cite{Cui1}.

\subsection{Measuring entanglement}
Entanglement measure quantifies how much entanglement is contained
in a quantum state. The entanglement measure for a state is zero iff
the state is separable, and the bigger is the entanglement measure,
then more entangled is the state. By axiomatic approach, (i) it is
any nonnegative real function of a state which can not increase
under local operations and classical communication (LOCC) (so called
monotonicity) \cite{Brandao}; (ii) it is zero for separable states;
(iii) and/or it satisfies normalization, asymptotic continuity, and
convexity. There are operational entanglement measures such as
distillable entanglement, distillable key and entanglement cost, as
well as abstractly defined measures such as ones based on convex
roof construction (e.g., concurrence and entanglement of formation)
or based on a distance from a set of separable states such as the
relative entropy of entanglement \cite{Horodecki}. One of the most
famous measures of entanglement is the Wootters' concurrence
\cite{Wootters} of two-qubit system. We will use it to study the
entanglement dynamics and obtain the entanglement transfers
\cite{Maruyama:07,Cui2,Cui3}.
 For a
system described by a density matrix $\rho$, the concurrence
$\mathcal{C}(\rho)$ is
\begin{equation}
\mathcal{C}(\rho)=\max(0,\sqrt{\lambda_1}-\sqrt{\lambda_2}-\sqrt{\lambda_3}-\sqrt{\lambda_4}),
\end{equation}
 where $\lambda_1, \lambda_2, \lambda_3$, and $\lambda_4$ are the
 eigenvalues (with $\lambda_1$ the largest one) of the ``spin-flipped"
 density operator $\zeta=\rho(\sigma_y^A\otimes\sigma_y^B)\rho^{*}(\sigma_y^A\otimes\sigma_y^B)$,
 where $\rho^{*}$ denotes the complex conjugate of $\rho$ and
 $\sigma_y$ is the Pauli matrix. $\mathcal{C}$ ranges in
 magnitude from 0 for a disentanglement state to 1 for a maximally entanglement state.

\begin{figure*}
\centerline{\scalebox{1.4}[0.9]{\includegraphics{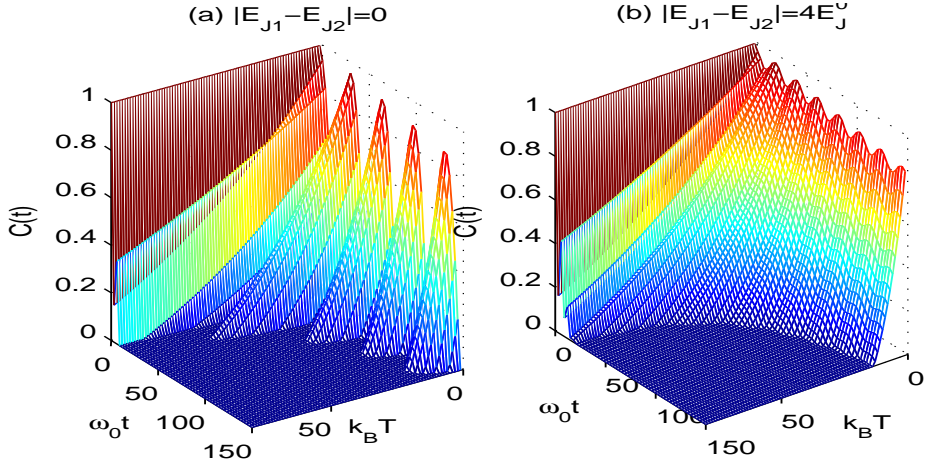}}}
\caption{Color online) Non-Markovian entanglement concurrence
$C_{\rho}(t)$ dynamics as a function of the thermal energy ``$k_BT$"
for the conditions $|E_{J1}-E_{J2}|=0$ and $|E_{J1}-E_{J2}|=4E_J^0$,
respectively.}
\end{figure*}

The initial state chosen is the previous generated maximal entangled
state,
$|\Psi\rangle=(|g\rangle_1|e\rangle_2-i|e\rangle_1|g\rangle_2)/\sqrt{2}$,
which is an ``X" form mixed state \cite{Yu1,Yu2,Yu3} that has
non-zero elements
 only along the main diagonal and anti-diagonal.
  The general
 form of an ``X" density matrix is as follows
\begin{equation}
\label{xstate}
 \rho=\left(\begin{array}{cccc}
a&0&0&m+in\\
0&b&e+if&0\\
0&e-if&c&0\\
m-in&0&0&d
\end{array}\right).
 \end{equation}
 Such states are general enough to include states such as the Werner
 states, the maximally entangled mixed states (MEMSs) and the
 Bell states; and it also arises in a wide variety of physical
 situations.
A remarkable aspect of the ``X" form mixed states is that the time
evolution of the master equation (5) determined by the initial ``X"
form is
 maintained during the evolution.
 This particular form of the density matrix allows us to
 analytically express the concurrence at time $t$ as
 \begin{equation}
 \mathcal{C}_{\rho}^{X}(t)=2\max\{0,K_1(t),K_2(t)\},
 \end{equation}
 where $K_1(t)=\sqrt{e^2(t)+f^2(t)}-\sqrt{a(t)d(t)}$, and
 $K_2(t)=\sqrt{m^2(t)+n^2(t)}-\sqrt{b(t)c(t)}$.

 The non-Markovian master equation (\ref{master equation}) is equivalent to a system of coupled differential
 equations, the first four of which describe the time evolution of
 the populations, namely
 \begin{eqnarray}
\dot{a}&=&-2(\Delta(t)+\gamma(t))a+(\Delta(t)-\gamma(t))(b+c)\nonumber\\
\dot{b}&=&(\Delta(t)+\gamma(t))a-2\Delta(t)b+(\Delta(t)-\gamma(t))d-2Jf\nonumber\\
\dot{c}&=&(\Delta(t)+\gamma(t))a-2\Delta(t)c+(\Delta(t)-\gamma(t))d+2Jf\nonumber\\
\dot{d}&=&(\Delta(t)+\gamma(t))(b+c)-2(\Delta(t)-\gamma(t))d,
 \end{eqnarray}
 while the other equations describe the time evolution of the
 coherence:
 \begin{eqnarray}
\dot{e}&=&-2\Delta(t)e+(E_{J1}-E_{J2})f\nonumber\\
\dot{f}&=&-2\Delta(t)f-(E_{J1}-E_{J2})e+J(b-c)\nonumber\\
\dot{m}&=&-2\Delta(t)m+(E_{J1}+E_{J2})n\nonumber\\
\dot{n}&=&-2\Delta(t)n-(E_{J1}+E_{J2})m.
 \end{eqnarray}
In the following subsection we will bring to light the features
characterizing the dynamics of superconducting entanglement.

\begin{figure*}
\centerline{\scalebox{0.8}[0.7]{\includegraphics{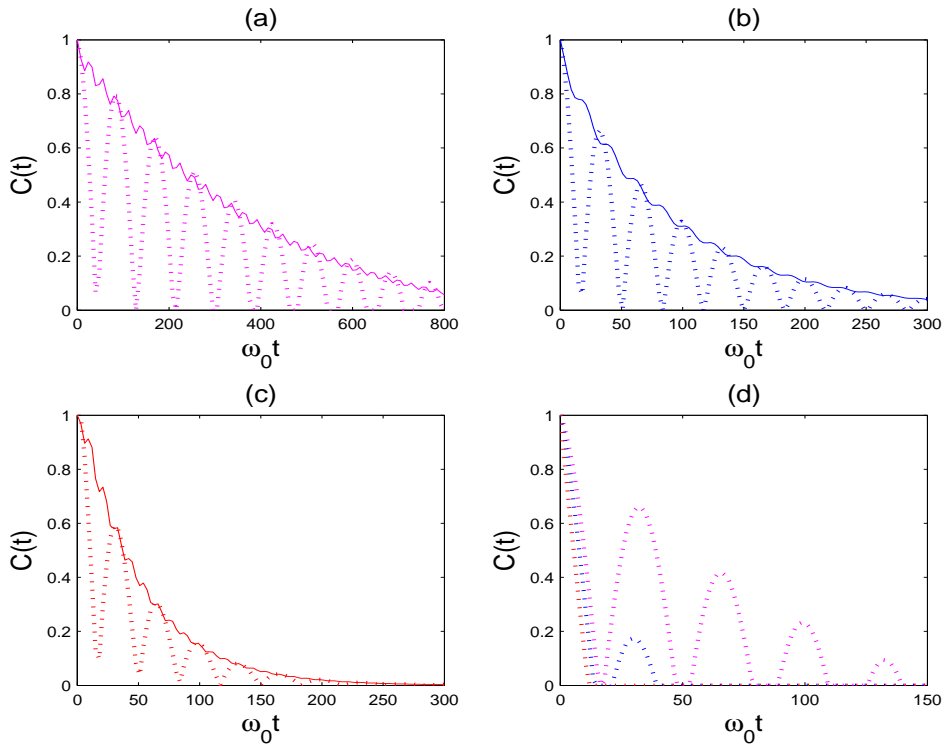}}}
\caption{(Color online) The evolutions of the concurrence
$\mathcal{C}(t)$ of the superconducting qubits in a low temperature
environment with $k_BT=0.03\omega_0$ for (a) ratio
$r\equiv\omega_c/\omega_0=0.3$, (b) $r=1$, and (c) $r=10$,
respectively. In these figures, we mark the solid line as
$|E_{J1}-E_{J2}|=4E_J^0$, and dashed line as $|E_{J1}-E_{J2}|=0$.
Figure (d): entanglement dynamics in a high temperature environment,
$k_BT= \omega_0$, magenta dashed line $r=0.3$,  blue dashed line
$r=1$, and red dashed line $r=10$. }
\end{figure*}
\subsection{Numerical demonstration}

Using the same experimental parameters as \cite{McDermott}:
$C_J=4.3pF, I_0=13.3\mu A$, $I_b=0.9725I_0$, $C_J/C\sim0.1$ and
$\omega_0=2\pi\times6$GHz, which is chosen as the norm unit. The
temperature is regarded as a key factor in a disentanglement
process. Another reservoir parameter playing a key role in the
entanglement dynamics  is the ratio $r=\omega_c/\omega_0$ between
the reservoir cutoff frequency $\omega_c$ and the system oscillator
frequency $\omega_0$. By varying these two parameters $k_BT$ and
$r$, the time evolution of the open system varies prominently for
different cases.

At first, let's consider two extreme cases, the strongest coupling
$(\Phi_k=0, \Phi_0)$, and the decoupling $(\Phi_k=\Phi_0/2)$, which
represent $E_{Jk}=\pm2E_J^0$ and $0$, respectively. In Fig. 2, the
time evolutions of the non-Markovian system concurrence for various
values of temperature are plotted in the two extreme cases
$|E_{J1}-E_{J2}|=0$ and $|E_{J1}-E_{J2}|=4E_J^0$. Here, we choose
the ratio $r=0.1$ and $k_BT$ ranges from $0$ to $100\omega_0$. We
can also compare the non-Markovian entanglement dynamics in the
above two cases clearly. Fig. 2(a) is the case of
$|E_{J1}-E_{J2}|=0$, which means both the two superconducting qubits
have the same phase $\Phi_k$. The oscillation of the concurrence is
also displayed in Fig. 2(a) besides the entanglement sudden death
(ESD) and entanglement sudden birth (ESB). The concurrence decays
with small amplitude in Fig. 2(b), $|E_{J1}-E_{J2}|=4E_J^0$.  Both
(a) and (b) show that the lower the temperature the more prominently
the entanglement. The sudden death and birth behaviors shown in
Fig.2 are new features for physical dissipation and is induced by
classical noise as well as quantum noise. The oscillating phenomenon
embodies the non-Markovian memory effect in Eq. (6). The concurrence
descended when $\Delta(t)-\gamma(t)>0$ and ascended when
$\Delta(t)-\gamma(t)<0$.

 In Fig. 3, we show entanglement dynamics
for (a) $r\ll1$ ( magenta line), (b) $r=1$ (blue line), and (c)
$r\gg1$ (red line) in a low temperature reservoir
($k_BT=0.03\omega_0$), which implies that the ratio
$r=\omega_c/\omega_0$ between the reservoir cutoff frequency
$\omega_c$ and the system oscillator frequency $\omega_0$ plays a
key role in the dynamics of the system. Obviously, the larger the
ratio $r$ the shorter the concurrence lasting time. Moreover, like
Fig. 2, the concurrence decays with small amplitude in the condition
$|E_{J1}-E_{J2}|=4E_J^0$ but has the same lasting time. The similar
results can be found in \cite{Nori:07,Nori:10}. In Fig. 3(d),
entanglement dynamics in high temperature environment, $k_BT=
\omega_0$, magenta dotted line $r=0.3$,  blue dotted line $r=1$, and
red dotted line $r=10$. In this figure we can see the ESD and ESB
phenomenons obviously.  From these simulations, we find that the
entanglement can be open-loop controlled and the ESD time can be
prolonged by adjusting temperature, $r$ and the superconducting
phases $\Phi_k$ in the superconducting qubit systems.

\section{Conclusions}
In the present work, we have theoretically studied the
 entanglement generation and dynamics in coupled superconducting qubits
 systems. We characterize the entanglement by the thermal energy $k_BT$,
 the ratio $r$ and the energy levels split of the superconducting
 circuit:
 $|E_{J1}(\Phi_1)-E_{J2}(\Phi_2)|$. Non-Markovian noise arising from
 the structured environment or from strong coupling appears to be
 more fundamental and the ESD and ESB phenomena are analyzed in this
 paper. Our simulation results demonstrated that  the lower the temperature the more prominent the
entanglement. Moreover, the ESB phenomenon embodies the
non-Markovian memory effect. We also find that the entanglement can
be open-loop controlled and the ESD time can be prolonged by
adjusting the temperature, $r$ and the superconducting phases
$\Phi_k$ in the superconducting qubit systems.
 Superconducting qubits offer evident advantages due
to their
 scalability and controllability. We hope that such techniques will
 be experimentally implemented in the near future.

\section*{Acknowledgments}
We thank the referee for useful suggestions and enlightening
comments. This work was supported by the National Natural Science
Foundation of China (No. 60774099, No. 60821091), the Chinese
Academy of Sciences (KJCX3-SYW-S01), and by the CAS Special Grant
for Postgraduate Research, Innovation and Practice.


\end{document}